\documentclass[aps,twocolumn,preprintnumbers,amsmath,amssymb]{revtex4}

\usepackage{graphicx}
\usepackage{dcolumn}
\usepackage{bm}
\usepackage{multirow}
\usepackage{color}
\usepackage{ulem}
\usepackage{amsmath}
\usepackage{gensymb}

\begin{document}
\title{Excitonic linewidth approaching the homogeneous limit in MoS$_2$ based \\ van der Waals heterostructures : accessing spin-valley dynamics}

\author{F. Cadiz$^{1}$}
\email{cadiz@insa-toulouse.fr}
\author{E. Courtade$^{1}$}
\author{C. Robert$^{1}$}
\author{G. Wang$^{1}$}
\author{Y. Shen$^{2}$}
\author{H. Cai$^{2}$}
\author{T. Taniguchi$^{3}$}
\author{K. Watanabe$^{3}$}
\author{H. Carrere$^{1}$}
\author{D. Lagarde$^{1}$}
\author{M. Manca$^{1}$}
\author{T. Amand$^{1}$}
\author{P. Renucci$^{1}$}
\author{S. Tongay$^{2}$}
\author{X. Marie$^{1}$}
\author{B. Urbaszek$^{1}$}
\email{urbaszek@insa-toulouse.fr}

\affiliation{
$^1$ Universit\'e de Toulouse, INSA-CNRS-UPS, LPCNO, 135 Av. Rangueil, 31077 Toulouse, France}
\affiliation{
$^2$ School for Engineering of Matter, Transport and Energy, Arizona State University, Tempe,AZ 85287,USA}
\affiliation{
$^3$ National Institute for Materials Science, Tsukuba, Ibaraki 305-0044, Japan}

\begin{abstract}
The strong light matter interaction and the valley selective optical selection rules make monolayer (ML) MoS$_2$ an exciting 2D material for fundamental physics and optoelectronics applications. But so far optical transition linewidths even at low temperature are typically as large as a few tens of meV and contain homogenous and inhomogeneous contributions. This prevented in-depth studies, in contrast to the better-characterized ML materials MoSe$_2$ and WSe$_2$. In this work we show that encapsulation of ML MoS$_2$ in hexagonal boron nitride can efficiently suppress the inhomogeneous contribution to the exciton linewidth, as we measure in photoluminescence and reflectivity a FWHM down to 2~meV at T = 4~K. This indicates that surface protection and substrate flatness are key ingredients for obtaining stable, high quality samples. Among the new possibilities offered by the well-defined optical transitions we measure the homogeneous broadening induced by the interaction with phonons in temperature dependent experiments. We uncover new information on spin and valley physics and present the rotation of valley coherence in applied magnetic fields perpendicular to the ML.   
\end{abstract}


\maketitle
\section{Introduction}
The first member of the transition metal dichalcogenides (TMDC) to be established as a direct gap semiconductor in monolayer (ML) form was MoS$_2$ \cite{Splendiani:2010a,Mak:2010a}, which has resulted in a global research effort exploring this promising 2D semiconductor family \cite{kim:2014a,Pospischil:2014a,Castellanos:2016a,Mak:2016a,Dufferwiel:2015a,Liu:2015a,Korn:2011a,Sidler:2016a,pollmann:2015a,Jakubczyk:2016a,Chernikov:2015b,Zeng:2012a,Yu:2016a,Neumann:2017a}. First proto-type device applications such as transistors \cite{Radisavljevic:2011a,Fang:2012a,Dery:2015a} and light emitters \cite{Sundaram:2013a,Ross:2014a,Zhang:2014b} have shown the promise of this atomically thin materials for electronics and optoelectronics. Another motivation for research into MoS$_2$ is the high natural abundance of the naturally occurring mineral molybdenite \cite{Dickinson:1923a}. Because of their apparent superior optical quality as compared to ML MoS$_2$, recent research on layered TMDC materials with emphasis on challenging experiments in valleytronics and light-matter interaction has focused mainly on the closely related monolayer materials WSe$_2$, WS$_2$ and MoSe$_2$ \cite{Mak:2016a,Xu:2014a}. In the transition metal \textit{diselenide} MLs, the full width at half maximum (FWHM) of the neutral exciton transition at the optical bandgap is typically of the order of 10 meV (50 meV)  at T=4~K (300~K) when exfoliated onto Si/SiO$_2$ substrates \cite{Ross:2013a,Jones:2013a,Wang:2015b}.\\
\indent Here we show that the optical transition linewidth in MoS$_2$ MLs encapsulated in hexagonal boron nitride (hBN) \cite{Taniguchi:2007a} reaches values down to 2~meV at T=4 K, see for example Figs.~3d and 2a. This is a major improvement on the usually reported emission linewidth of $\approx 50$~meV at low temperature \cite{Korn:2011a,Kioseoglou:2012a,Cao:2012a,Lagarde:2014a,Zeng:2012a,Sallen:2011a,Stier:2016a,Mitioglu:2016a,Neumann:2017a}. The high quality van der Waals heterostructures \cite{Geim:2013a} investigated here allow accessing new information on their optical and spin-valley properties :\\
\begin{figure*}
\includegraphics[width=0.9\textwidth]{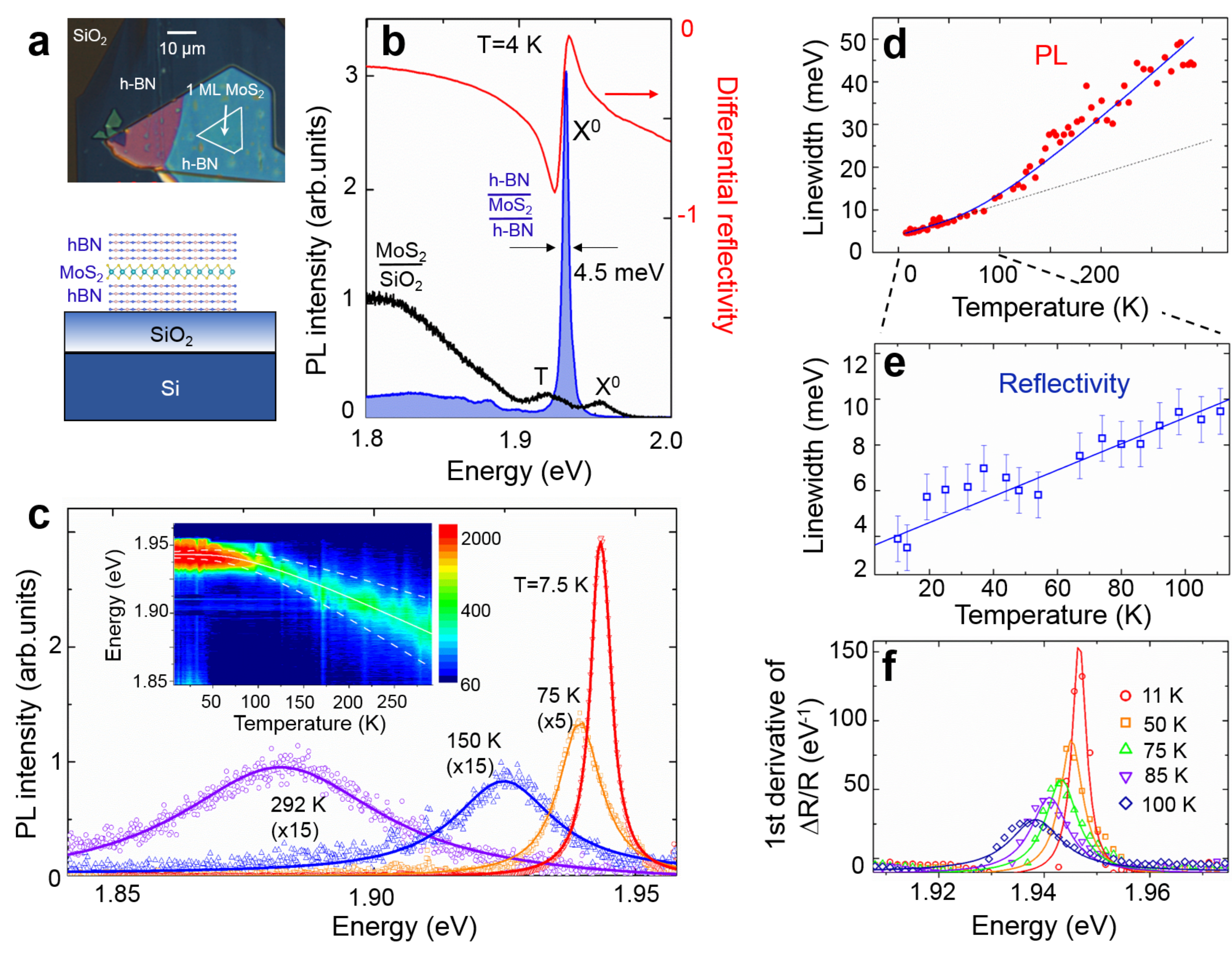}
\caption{\label{fig:fig1} (a) Top: Optical microscope image of van der Waals heterostructure hBN / ML MoS$_2$ / hBN. Bottom: schematic of sample (b)  PL spectrum (filled curve) at T$=4$ K for one capped ML when excited with a $2.33$ eV cw laser with a power density of $50\;\mu$W/ $\mu \mbox{m}^2$.  The neutral exciton transition, which is also the main feature of the differential reflectivity spectrum (red curve), exhibits a very narrow linewidth, of 4.5 meV for this particular sample. Also shown is the PL spectrum of an uncapped  MoS$_2$ ML deposited directly onto the SiO$_2$ substrate (black curve) measured under the same conditions. (c)  PL spectrum of a capped sample for selected sample temperatures. The inset shows a colormap revealing the temperature-evolution of the spectrum's intensity. The white, full line tracks the evolution of the peak position according to Eq.(\ref{eq1}), whereas the dashed lines are a guide to the eye indicating the linewidth (FWHM). The excitation density was kept as low as $1\;\mu$W/$\mu \mbox{m}^2$.
(d) Temperature evolution of the neutral exciton linewidth extracted from the PL spectra when excited with $1\;\mu$W/$\mu \mbox{m}^2$.Typical error bars are $\pm1$~meV ($\pm5$~meV) at 4~K (300~K). The solid line is a fit according to Eq.(\ref{eq2}), and the dashed line represents the linear term which dominates at low temperatures.(e) Linewidth extracted from the first derivative of the differential reflectivity as a function of temperature. The linear fit is consistent with a phonon-induced broadening. (f) The first derivative of the differential reflectivity for selected temperatures, centered around the X$^0$ absorption. 
}
\end{figure*}
\indent (1) The FWHM of the transition linewidth measured in photoluminescence (PL) and reflectivity down to 2~meV provides a new upper limit for the homogeneous linewidth of the exciton transition, corresponding to a lower limit for exciton radiative lifetime of $T_1 \sim 330$~fs. 
The linewidth of $2$ meV is even smaller than a previously reported value of the homogeneous linewidth in ML MoS$_2$ on SiO$_2$ substrates (4.5~meV) measured with four wave mixing \cite{Dey:2016a}. In temperature dependent experiments we measure the broadening of the optical transition linewidth due to interactions with acoustic and optical phonons. The proximity of the $\Gamma$-point close in energy to the valence band maximum at the $K$-point in MoS$_2$ \cite{Cheiwchanchamnangij:2012a,Kormanyos:2013a, Yuan:2016a} might be at the origin of the strong broadening observed when approaching room temperature. \\ 
\indent (2) Remarkably, the emission linewidth and intensity in our hBN / ML MoS$_2$ / hBN structures are unchanged following several cool-down cycles to T=4~K and exposure to laser radiation. 
This is in contrast to the variable optical response of uncapped MoS$_2$ \cite{Cadiz:2016a} and WS$_2$ \cite{Currie:2015a,He:2016a} MLs exfoliated onto Si/SiO$_2$ substrates, which often depends also on cool-down procedure and sample history. Reproducibility and stability are essential to better understand the unusual optical properties of these materials \cite{Xu:2014a,Mak:2016a} and to prepare possible device applications. We observe these narrow transition lines in different ML MoS$_2$ samples exfoliated from different bulk material (i) grown by chemical vapor transport (CVT, see sec.\ref{apendix}) and (ii) commercially available crystals.  \\
\indent (3) The narrow linewidth allows us to distinguish spectrally the emission from the non-identical valleys $K^+$ and $K^-$ as they are split by the valley Zeeman effect \cite{Li:2014a,srivastava:2015,Macneill:2015a,Aivazian:2015a,Arora:2016a,Wang:2015d}. We are able to determine the longitudinal exciton Land\'e g-factor in magnetic fields below 10~T in a commercial magneto-cryostat. Due to the large linewidth of uncapped samples, this measurement was so far only possible in pulsed field facilities generating several tens of Tesla \cite{Stier:2016a,Mitioglu:2016a}. We find for the neutral exciton $g_X=-1.7 \pm 0.1$, different from earlier reports on unprotected samples that suggested $g_X=-4$. Here we discuss the possible impact of doping (i.e. contributions from charged excitons) and the dielectric environment. We also show the \textit{generation} and \textit{rotation} of robust valley coherence, a coherent superposition of valley states using the chiral optical selection rules \cite{Jones:2013a}, an important step towards full optical control of valley states in these 2D materials \cite{Schmidt:2016a,Wang:2016c,Ye:2017a}.
\section{Neutral exciton transition in PL and reflectivity}
\subsection{Sharp linewidth in hBN / ML MoS$_2$ / hBN}
The photoluminescence (PL) spectrum at low temperature of MoS$_2$ MLs encapsulated in hBN  exhibits a very narrow neutral exciton emission (X$^0$) at energies between $1.93-1.95$ eV with a linewidth (FWHM) varying between $2$ and $5$ meV depending on the sample used and the detection spot position, where the detection spot diameter is $\approx 1~\mu$m (see Sect.~V) . The low-temperature linewidth in our capped samples is more than one order of magnitude smaller than the one observed for the usual broad peak reported in most studies \cite{Lagarde:2014a,Zeng:2012a,Sallen:2011a,Korn:2011a,Cao:2012a,Stier:2016a,Mitioglu:2016a,Kioseoglou:2012a,Neumann:2017a}. This  transition usually attributed to the neutral exciton, is possibly merged with a broad charged exciton and/or defect contribution, as discussed recently \cite{Cadiz:2016a}. The PL and reflectivity FWHM measured here is for certain samples smaller than the  homogeneous linewidth extracted by means of four wave mixing techniques for MLs on SiO$_2$ \cite{Dey:2016a}. For comparison, PL and reflectivity give neutral exciton linewidth down to 4~meV in hBN encapsulated WSe$_2$ MLs  \cite{Manca:2016a, Chow:2017a,Jin:2016a}. \\
\indent A typical PL emission spectrum of a capped sample is shown in Fig.\ref{fig:fig1}b (filled curve) under a 50 $\mu$W cw laser excitation at 2.33 eV. The neutral exciton emission, identified by polarization-resolved experiments (shown later in sec.\ref{valleypol}), corresponds to the main feature observed in the differential reflectivity spectrum in the same figure.\\
\indent Note that almost no signature of charged excitons or defect-related emission is observed in these MLs. In contrast, the typical spectrum of  a MoS$_2$ ML  deposited directly onto the SiO$_2$ substrate (black curve) exhibits a very large and broad defect-related emission followed by charged exciton (trion) emission (FWHM 38 meV) and a neutral exciton emission (FWHM 16 meV) \cite{cadiz:2016b} whose intensity drops dramatically after a few minutes of laser exposure due to laser-induced doping of these MLs  \cite{Cadiz:2016a} (here, the spectrum was recorded within the first 0.5 seconds following laser exposure).  The intensity stemming from the neutral exciton in our hBN-protected samples is at least one order of magnitude higher than in unprotected ones, for a fully quantitative comparison cavity effects need to be taken into account \cite{Lien:2015a}. \\
\indent Important for reproducibility, no evolution or hysteresis of the PL spectrum was observed even after several minutes of exposure to $\sim 500$ $\mu$W excitation and after several temperature cycles $4\leftrightarrow300$~K. These results demonstrate that capping the MLs with hBN allows to address the intrinsic high optical quality of these 2D crystals, and protects them from possible charge transfers and local electric field fluctuations coming from the substrate. The layer of hBN below MoS$_2$ acts as a spacer so that the surface roughness of SiO$_2$ is not transferred to the ML \cite{Man:2016a}. This also suggests that for certain applications no acid treatment of the surface is needed if the freshly exfoliated sample is capped immediately with hBN \cite{Amani:2015a}. Note that the neutral exciton X$^0$ line is redshifted by 20 to 30~meV in capped samples with respect to unprotected ones (typically found at $\sim 1.96-1.97$ eV  \cite{cadiz:2016b}). This was also observed in capped WSe$_2$ MLs \cite{Manca:2016a, Chow:2017a,Jin:2016a}, and may be due to a different dielectric environment when the ML is capped with hBN \cite{Chernikov:2014a,Stier:2016b} or to a different strain when the MLs are not directly in contact with  the SiO$_2$ substrate. In recent studies addressing this problem in ML WSe$_2$ it has been suggested that this small red shift of the emission is the result of two substantial shifts compensating each other \cite{Stier:2016b}: the exciton binding energy $E_b$ considerably decreases, but the free carrier bandgap $E_g$ as well, which results in a very small shift of the optical bandgap $E_g-E_B$, which is the neutral exciton emission energy.
\subsection{Temperature induced line broadening}
Figure \ref{fig:fig1}c shows the PL spectrum of an encapsulated sample for selected temperatures, revealing a blueshift of $62$ meV and a narrowing by a factor of 10 for the neutral exciton emission when cooling down the sample from 300 to 4~K.
The inset is a colormap showing the evolution of the PL  intensity as a function of temperature for a fixed low laser power of $1\;\mu$W at $1.96$ eV. When going from room temperature to $4$~K, the intensity increases significantly and the linewidth narrows down to only a few meV.  In addition to the reduction of non-radiative recombination on defects, the overall, not monotonous increase in intensity when reaching low temperature may be due to a competition between bright and dark excitonic states and possibly indicates that ML MoS$_2$ belongs to the bright family as ML MoSe$_2$. Several theoretical calculations have predicted that the lowest excitonic transition in MoS$_2$ MLs should be bright \cite{Kormanyos:2015a,Echeverry:2016a}, but this is still debated since the opposite configuration has also been predicted \cite{Qiu:2015a}. Here charge tunable samples \cite{Wang:2017a} or experiments in transverse magnetic fields \cite{Zhang:2017a,Molas:2017a} are expected to give a better indication on the dark-bright order in future experiments on these narrow linewidth samples.\\
The temperature dependence of the PL peak position follows a standard hyperbolic cotangent relation \cite{Odonnel:1991a}:
\begin{equation}
E_G(T)=E_G(0) - S \langle  \hbar w \rangle [\coth (\langle \hbar w\rangle/(2k_BT)  ) -1]
\label{eq1}
\end{equation}
where $E_G(0)$ is the optical bandgap at zero temperature, $S$ is a dimensionless coupling constant, $k_B$ is Boltzmann's constant and $\langle \hbar w\rangle$ is an average phonon energy. The fitted curve (solid line shown in the inset of Fig.\ref{fig:fig1}c ) yields $E_G(0)=1.943$ eV, $S=1.87$ and $\langle \hbar w\rangle= 24.2\pm1.5$~meV. The coupling constant and the average phonon energy are similar to those reported previously in the literature for the TMDC ML family \cite{Mitioglu:2016a,Christopher:2016a,Ross:2013a,Currie:2015a,Kioseoglou:2016b}. The evolution of the linewidth as a function of temperature is shown in panel d of Fig.\ref{fig:fig1}, and it can be seen that it decreases by more than one order of magnitude when passing from room to cryogenic temperatures, which is a signature of an efficient coupling with phonons. The evolution of the linewidth can be phenomenologically approximated by a phonon-induced broadening \cite{Selig:2016a,Dey:2016a} :
\begin{equation} \gamma= \gamma_0 + c_1 T +  \frac{c_2}{e^{\Omega/k_BT}-1} \label{eq2} \end{equation}
where $\gamma_0=4 \pm 0.2$~meV and $c_1=70\pm 5 \mu$eV/$\mbox{K}^{-1}$ describes the linear increase due to acoustic phonons, and is of the same order of magnitude than the one reported in Refs.\cite{Dey:2016a,Moody:2015a}. The parameter $c_2=42.6 \pm 1.2$~meV is a measure of the strength of the phonon-coupling and $\Omega=24.2$~meV is the \textit{averaged} energy of the relevant phonons, that we obtained by fitting the optical bandgap energy shift with Eq.(\ref{eq1}) for consistency. Note that the value of $c_2=42.6$ meV found here is larger than the $6.5$ meV and $15.6$ meV previously reported for WS$_2$ and MoSe$_2$ MLs, respectively \cite{Selig:2016a}, but so far no detailed discussion exists on MoS$_2$ MLs since the optical transition linewidth in uncapped samples is too broad. This larger coupling with optical phonons may be a consequence of the energy-proximity between the local maximum of the valence band at $\Gamma$ and the absolute maxima at the $K$ valleys in MoS$_2$ MLs \cite{Cheiwchanchamnangij:2012a,Kormanyos:2013a, Yuan:2016a}.\\
\indent In order to exclude any possible problems due to localization effects on the linewidth evolution at low temperatures, we have performed reflectivity measurements in addition to PL as a function of temperature. The first derivative of the differential reflectivity for selected temperatures is shown in Fig.\ref{fig:fig1}f, and also exhibits a very small linewidth of $\sim 3$ meV which broadens roughly linearly as a function of temperature up to 100 K, as shown in Fig.\ref{fig:fig1}e. A linear fit gives a slope of $61 \pm 5 mu$eV/K, similar to the $70\pm 5 \mu$eV/K observed for the PL linewidth.  Since reflectivity is sensitive to transitions with a large density of states \cite{Klingshirn:2007} i.e. free excitons, we conclude from the very similar temperature dependence that the PL also stems dominantly from free neutral excitons. Due to sample inhomogeneity the exciton energy fluctuates by a few meV for different points on the sample, we found a maximum energy difference for the exciton PL energy of 10 meV between two extreme positions on the same flake \cite{StokesML}. 
\begin{figure}
\includegraphics[width=0.45\textwidth]{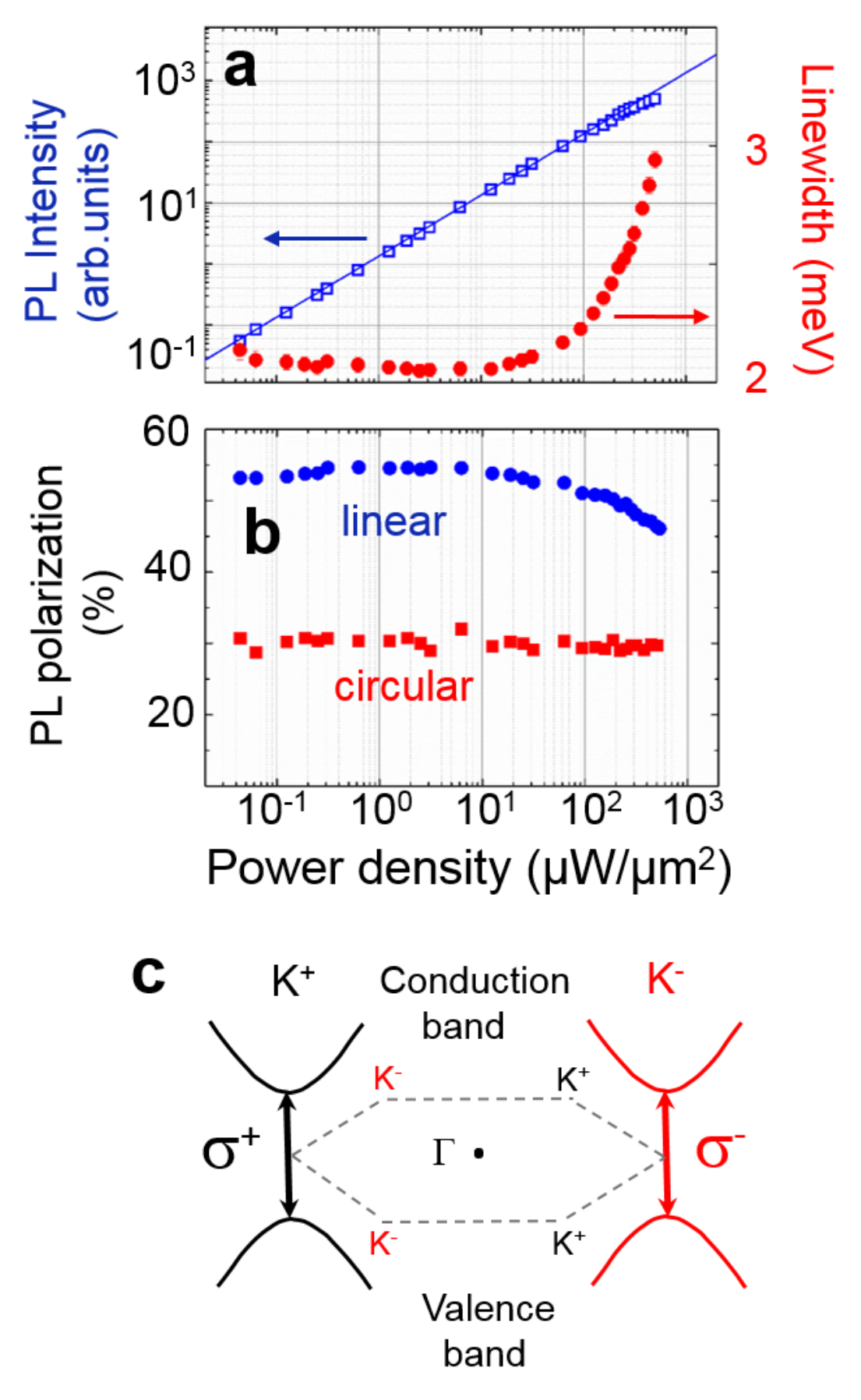}
\caption{\label{fig:fig2}  T=4~K. hBN / ML MoS$_2$ / hBN. (a) Integrated PL intensity (open squares) and linewidth  (full circles) as a function of excitation power density. The full line represents a  linear relationship between intensity and power density. 
(b) Degree of linear PL polarization $P_{\mbox{lin}}=(I^{X}-I^{Y})/(I^{X}+I^{Y})$ and circular polarization $P_c=(I^{\sigma^+}-I^{\sigma^-}  )/(I^{\sigma^+}+I^{\sigma^-})$, for linearly and circularly polarized excitation, respectively, as a function of laser power density. Laser energy 1.96~eV.
(c) chiral optical selection rules for interband optical transitions in non-equivalent valleys $K^+$ and $K^-$ after \cite{Xiao:2012a,Cao:2012a}}
\end{figure}
\section{Valley polarization and valley coherence manipulation}\label{valleypol}
\subsection{Measurements without applied magnetic fields}
The strong excitonic resonances with sharp linewidth in our sample make it possible to study optically the valley polarization in ML MoS$_2$ in great detail. Here we benefit from the chiral optical selection rules in TMDC monolayers, that allow optical excitation in the $K^+$ or $K^-$ valley, using $\sigma^+$ or $\sigma^-$ polarized laser excitation, respectively \cite{Xiao:2012a,Cao:2012a,Mak:2012a,Sallen:2012a,Jones:2013a}, see scheme in Fig.~\ref{fig:fig2}c. Detailed studies of valley polarization and coherence have  so far mainly focussed on ML WSe$_2$, due to the comparatively narrow linewidth and high polarization degree of the emission \cite{Jones:2013a,Hao:2015a}. Our results below show that the optical quality of ML MoS$_2$ is high enough to measure the Land\'e g-factor with high accuracy and study valley coherence rotation. In this section all experiments are carried out at T=4~K.\\
\begin{figure*}
\includegraphics[width=0.95\textwidth]{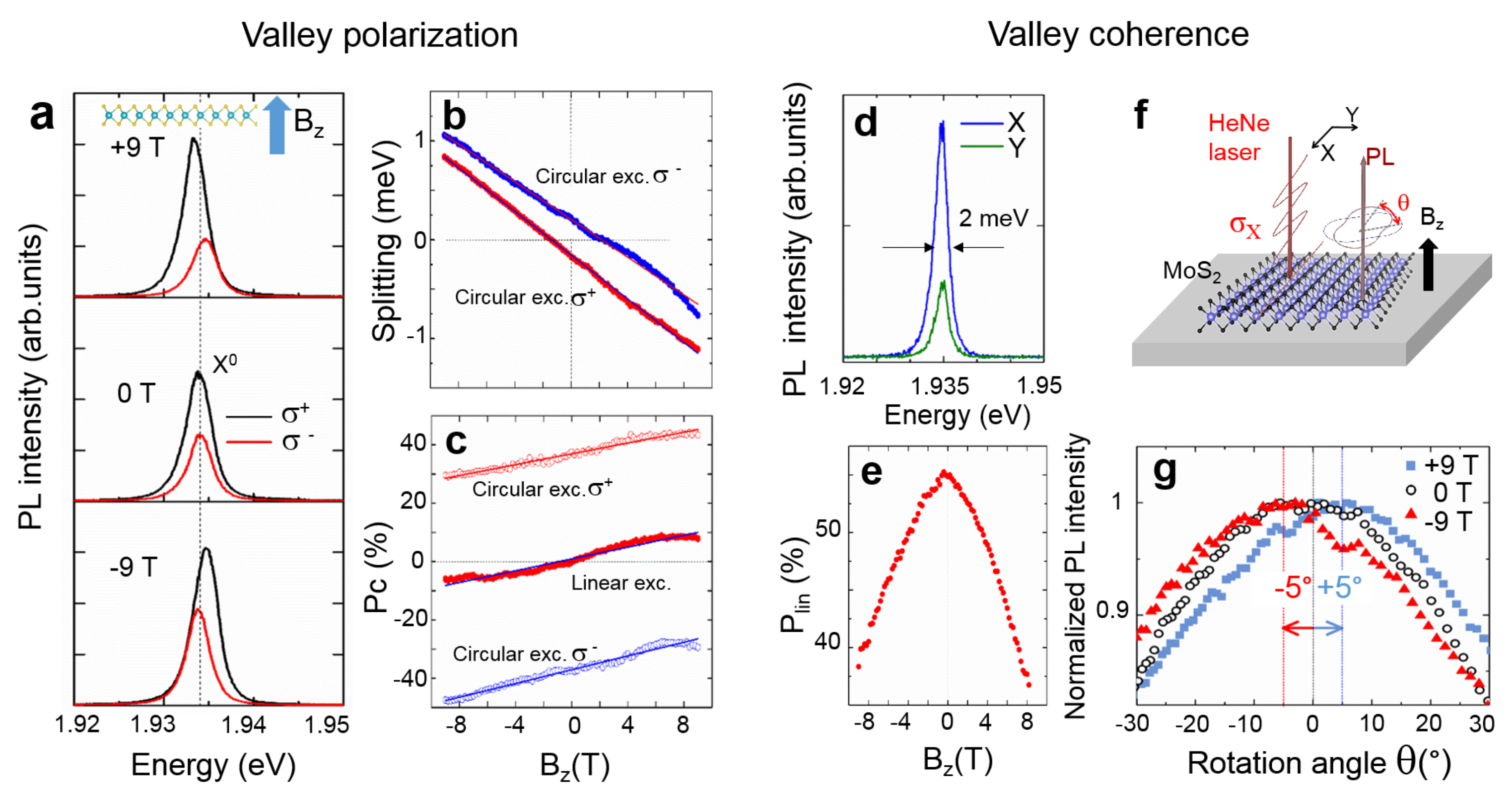}
\caption{\label{fig:fig3}  T=4~K. hBN / ML MoS$_2$ / hBN. (a) The left panel shows typical PL spectra for a capped MoS$_2$ ML under $\sigma^+$ circularly polarized excitation at selected values of the applied magnetic field $B_z$. Both circularly-polarized components of the PL are shown. (b) Zeeman splitting between both circular components of the PL for $\sigma^+$ and $\sigma^-$ polarized excitation at $10\;\mu\mbox{W}/\mu \mbox{m}^2$.  The measured splitting corresponds to an exciton g-factor of $g_X=-1.7 \pm 0.1$. (c) Degree of circular polarization $P_c$  as a function of the magnetic field and for different incident polarization of the excitation laser. (d) Degree of linear polarization $P_{\mbox{lin}}$  following linear excitation as a function of the magnetic field. e) Linearly polarized components of the PL spectrum after linearly polarized excitation along the X direction at 633 nm. The X$^0$ emission is highly co-linearly polarized, revealing robust valley coherence. f) Schematic of the experimental configuration used to control the valley coherence. (g) PL intensity under linear excitation as a function of the angle of the linear polarization analyzer, for selected values of the magnetic field. A rotation of $\pm 5\;\degree$ is observed for the PL polarization at $B_z=\pm 9$ T. 
}
\end{figure*}
\indent Excitation of MoS$_2$ MLs with a cw, circularly $\sigma^+$ polarized laser results in strongly  $\sigma^+$ polarized PL of the neutral exciton X$^0$, as shown in Fig.~\ref{fig:fig2}b (see also the spectra in Fig.~\ref{fig:fig3}a).We measure a high circular polarization degree of the PL of $P_c \sim 30\;\%$ corresponding to initialization of valley polarization. Remarkably the measured value of $P_c$ it is very robust and is independent of the explored laser excitation density which spans 4 orders of magnitude. \\
\indent We have also explored the generation of robust valley coherence \cite{Jones:2013a} in these capped MLs. As circularly polarized laser excitation induces valley polarization, excitation with a linearly polarized laser can generate a coherent superposition of valley states in the ML, referred to as valley coherence or optical alignment of excitons \cite{Meier:1984a}. As shown in panel b) of Fig.\ref{fig:fig2}, excitation with a linearly polarized laser results in a highly linearly polarized PL emission for the neutral exciton, with a degree of linear polarization as high as $P_{\mbox{lin}} \sim 55 \; \%$ in steady state (the polarization-resolved components of the PL are shown in Fig.\ref{fig:fig3}d).  Obtaining higher $P_{\mbox{lin}}$ than $P_c$ can be explained by the fact that the upper limit for the coherence time $T_2$ is twice the polarization lifetime $T_1$, which can result in higher $P_{\mbox{lin}}$ in steady state PL \cite{Dyakonov:2008a}. Here also the effects of the Coulomb exchange interaction on the polarization eigenstates can play a role \cite{Glazov:2014a}. 
Note that the valley coherence starts to drop for excitation densities above $50\;\mu\mbox{W}/\mu\mbox{m}^2$, probably due to the effect of exciton-exciton interactions on valley coherence \cite{Jeune:1998a}. The X$^0$ integrated intensity depends linearly on the excitation density, as shown in Fig.\ref{fig:fig2}a, and  the linewidth increases rapidly for laser powers above $100 \;\mu\mbox{W}/\mu\mbox{m}^2$. This broadening is a signature of an excitation-induced dephasing that may be also responsible for the observed drop of the valley coherence. Note that the absence of charged exciton emission in PL and in reflectivity suggests that this dephasing is dominated by exciton-exciton  interactions, and not by exciton-carrier collisions.
\subsection{Valley Zeeman splitting and field induced valley polarization}
\begin{figure}
\includegraphics[width=0.45\textwidth]{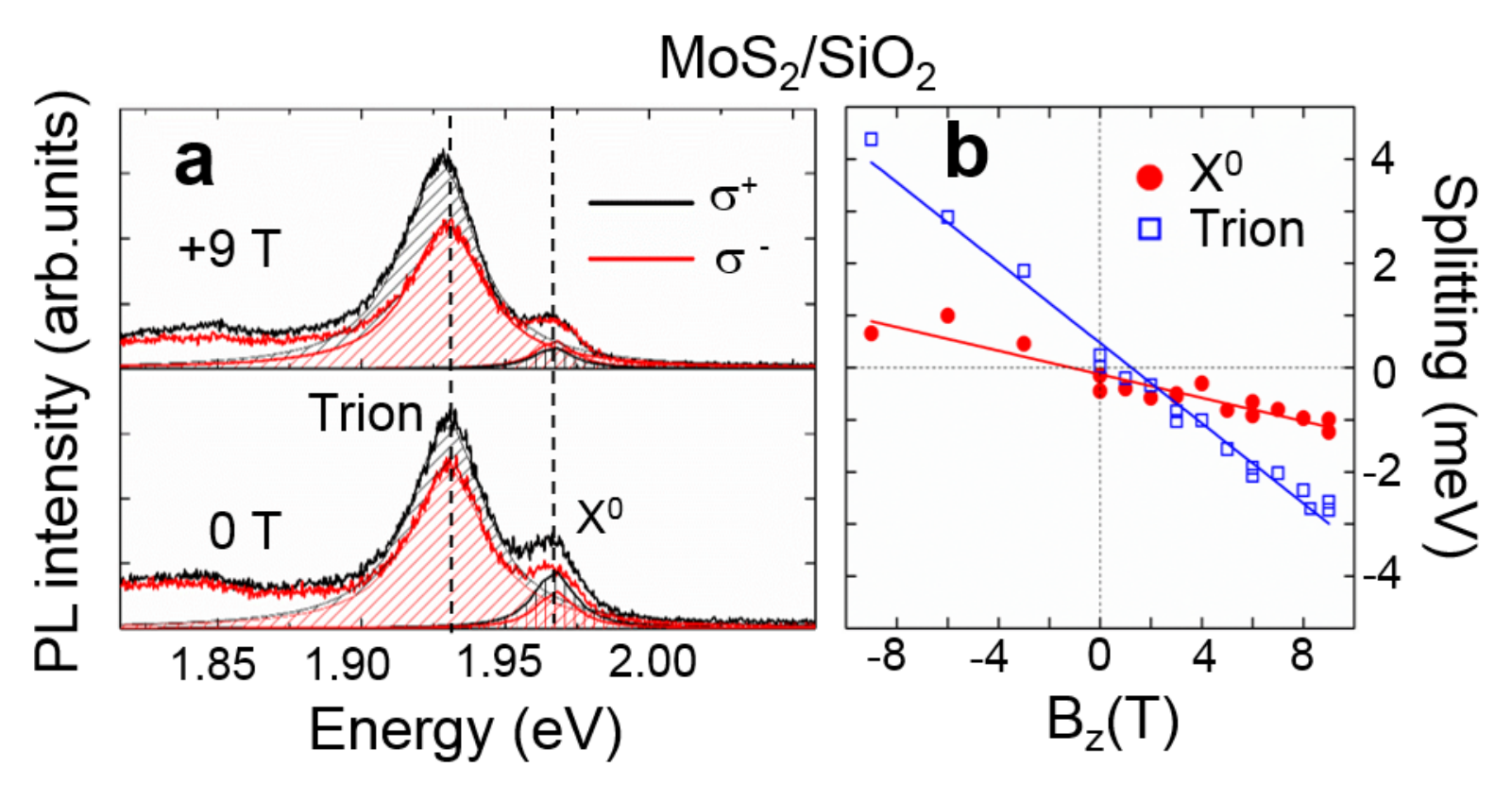}
\caption{\label{fig:fig4}  T=4~K.  ML MoS$_2$ / SiO$_2$/Si. (a) Typical PL spectra for an acid-treated, uncapped  MoS$_2$ ML at selected values of the applied magnetic field. A multiple-Lorentzian fit is employed in order to extract the Zeeman splitting of both neutral and charged exciton (trion) lines, shown in panel (b). Valley Zeeman splitting as a function of the applied field $B_z$. Land\'e g-factors of $g_X= -1.9 \pm 0.2$ and $g_T= -6.6 \pm 0.2$ are found. 
}
\end{figure}

Measuring the Land\'e g-factor is important for spin-valley physics, as it determines the energy separation achievable between the different polarization states \cite{Li:2014a,srivastava:2015,Macneill:2015a,Aivazian:2015a,Arora:2016a,Wang:2015d}. The g-factor, closely related to the effective mass tensor, also gives a fingerprint of the impact of different bands on the optical transitions \cite{Ivchenko:2005a,Macneill:2015a,Wang:2015d}. In addition spectacular effects are expected for monolayers with tunable electron density, where the evolution of the valley polarization and the valley Zeeman splitting has been interpreted in terms of a Fermi-polaron model for excitonic transitions, i.e. the simple definitions of neutral exciton and trion are replaced by the attractive and repulsive polaron \cite{Sidler:2016a,Efimkin:2017a,Back:2017a}.\\
\indent Thanks to the very narrow exciton emission linewidth in our protected samples, we were able to perform magneto-PL experiments and observe a clear valley Zeeman splitting at moderate magnetic fields. As shown in the schematics of Fig.\ref{fig:fig3}a, a magnetic field up to $\pm 9$ T is applied perpendicularly to the ML plane and the circular components of the PL are recorded as a function of the magnetic field. Figure \ref{fig:fig3}a shows the polarization-resolved PL spectra at T$=4$ K for different values of the magnetic field using a circularly-polarized cw laser at $1.96$ eV with a $10\;\mu\mbox{W}/\mbox{cm}^2$ power density. Note that the linewidth is comparable to the typical valley Zeeman splitting $\Delta_z$ at $10$ T, defined as the shift between the $\sigma^+$ and $\sigma^-$ polarized components of the PL. This energy splitting is found to depend linearly on the applied magnetic field (as shown in Fig.\ref{fig:fig3}b), and using $\Delta_z = g_X \mu_B B_Z$ where $\mu_B$ is the Bohr magneton, we extract an exciton g-factor of $g_{X}= -1.7 \pm 0.1$. Reducing the incident power by 4 orders of magnitude did not result in any measurable change in the extracted g-factor.\\
\indent In addition we note that the circular polarization of the PL  can be varied with the magnetic field, as shown in Fig.\ref{fig:fig3}c. At high positive magnetic fields, the valley splitting shifts the $K^+$ valley to lower energies, favouring the population of this valley in steady state conditions. In contrast, at high negative magnetic fields, populating the $K^+$ valley is energetically unfavourable, and as a consequence there is a reduction of the steady-state valley polarization. Independent of the incident polarization of the laser, we observe an increase of the valley polarization at a rate of $1\;\%/\mbox{T}$. Note that in addition to simple arguments on population also coupling to different exciton states, such as optically dark states, as function of magnetic field can have an impact on the observed emission intensities in the circular basis. \\
\indent In the literature, only a few reported experimental values exist for the exciton g-factor on MoS$_2$ MLs \cite{Stier:2016a,Mitioglu:2016a} between $g_X=-4$ and $g_X=-4.5$, obtained with magneto-reflectivity measurements on CVD-grown MoS$_2$ MLs.  In order to rule out any possible modification of the g-factor induced by the hBN encapsulation, we have performed magneto-PL  on uncapped MoS$_2$ MLs deposited directly onto SiO$_2$ substrates. In order to be able to identify the neutral exciton transition, the sample was first treated with  bis(trifluoromethane)sulfonimide (TFSI) \cite{Amani:2015a, cadiz:2016b} so that the emission coming from defects is strongly reduced. In order to minimize the impact of laser-induced doping of the ML and to avoid a complete disappearance of the X$^0$ feature in the PL spectra,  a maximum laser power of
$30\;\mu$W was used \cite{Cadiz:2016a}. Typical spectra are shown in panel a of Fig.\ref{fig:fig4}, dominated by a  trion peak of 32 meV linewidth and a smaller neutral exciton peak of 16 meV linewidth. The extracted Zeeman splitting for the neutral exciton, shown in panel b of Fig.\ref{fig:fig4}, is consistent with a g factor of $g_X=- 1.95 \pm 0.2$, close to the value obtained in capped MLs (Fig.~\ref{fig:fig3}). 
For the trion, we find a larger g-factor of $g_T= -6.6 \pm 0.2$. Note that we cannot exclude a contribution of defect-related emission to the measured g-factor of the trion peak due to its larger linewidth. 
Our measurements suggest that the neutral exciton g-factor of MoS$_2$ MLs might be intrinsically smaller (in magnitude) than for diselenide MLs and is different from the commonly observed value $g_X\approx -4$ \cite{Li:2014a,srivastava:2015,Macneill:2015a,Arora:2016a,Wang:2015d}. Variations in the measured neutral exciton g-factor can have different reasons: (i) different strain in CVD-grown MLs  as compared to encapsulated monolayers may play a role, (ii) Comparing g-factors in samples of different origin with different electron concentrations is difficult due to the possible impact of many-body interactions \cite{Sidler:2016a,Efimkin:2017a,Back:2017a}, (iii) eventually in unprotected samples with considerably broader linewidth, magneto-reflectivity may provide a g-factor that is a average of exciton and trion g-factors.
\subsection{Rotation of valley coherence}
In addition to the optical generation of valley \textit{polarization}, also the optical generation of valley \textit{coherence} has so far been mainly studied in WSe$_2$ and WS$_2$ and not MoS$_2$ due to the broad transition linewidth. Generation of valley coherence is achieved optically by linearly polarized excitation $\sigma_{X}$, which results in strongly linearly polarized neutral exciton (X$^0$) emission \cite{Jones:2013a}. Recent experiments have shown that this valley polarization can be rotated in ML WSe$_2$ and WS$_2$ \cite{Wang:2016c,Schmidt:2016a,Ye:2017a}, a first important step towards full control of valley states \cite{Press:2008a}. The MoS$_2$ ML is excited by a linearly polarized ($\sigma_X$) continuous wave He-Ne laser (1.96~eV) to generate valley coherence (i.e. optical alignment of excitons \cite{Meier:1984a}). Our target is to detect the neutral exciton X$^0$ valley coherence in the linear basis in PL emission. A liquid crystal based linear polarization rotator is applied in the detection path, to detect a possible rotation of the linear basis of the PL signal with respect to the initial linear excitation basis given by the laser. The rotation angle $\theta$ can be tuned by varying $B_z$. This approach avoids any macroscopic mechanical movement during the measurement and gives an accurate map of the angle dependent PL intensity as schematically illustrated in Fig.~\ref{fig:fig3}f, the full data set is plotted in Fig.~\ref{fig:fig3}g. The initial linear polarization corresponds to $\theta=0^\circ$ and we rotate the linear polarizer in detection and record for which angle $\theta$ the PL emission is maximised. \\
\indent In this experiment the external magnetic field lifts the valley degeneracy \cite{Li:2014a,srivastava:2015,Macneill:2015a,Aivazian:2015a,Arora:2016a,Wang:2015d} and results in a change of the oscillation frequency of the coherent superposition of valley states. This corresponds to a rotation of valley coherence (i.e. the exciton pseudo-spin) and we clearly measure this rotation in our experiments with angles up to $\pm 5$ degrees at $B=\pm 9$~T. As the rotation is measurable in our simple steady state experiment, we can conclude that the valley coherence time of the neutral excitons is at least roughly of the order of the PL emission time in our sample.\\
\indent A different measurement is presented in Fig.\ref{fig:fig3}e : we show the evolution of the linear polarization following linear excitation as a function of the magnetic field, but keeping the linear basis in detection fixed. A significant decrease is observed, which can have several origins: First, strong rotation of valley coherence resulting in small projection in the initial basis. As we measure only a small rotation angle $\theta$, this rotation can only be partially responsible for the observed drop of the linear polarization. A second cause for the decrease in the detected linear polarization can be dephasing induced during the PL lifetime \cite{Hao:2015a}. Note that a smaller rotation angle of the PL with respect to the one observed in WSe$_2$\cite{Wang:2016c} and WS$_2$ \cite{Schmidt:2016a} MLs (characterized by larger g-factors) is indeed expected for a smaller exciton g-factor measured in our sample. \\
\section{Conclusion and Perspectives}
In conclusion, we have shown that encapsulating MoS$_2$ MLs in hBN allows to access the intrinsic high optical quality of these 2D crystals. Key roles of the top and bottom hBN layers are the protection of the sample surface from possible physio- and chemiosorption during the experiment, providing an atomically flat surface for sample exfoliation to avoid ripples \cite{Man:2016a} and preventing optical doping from the SiO$_2$. The observed linewidth of a few meV at low temperatures is the smallest ever reported for this material  and provides an upper limit for the homogeneous linewidth ($\leq 2$ meV) and a lower limit for exciton radiative lifetime ($\geq 0.33$ ps). This confirms that the large PL linewidth of tens of meV reported for uncapped MoS$_2$ on SiO$_2$ is mainly caused by inhomogeneous broadening \cite{Dey:2016a}. No hysteresis of the PL after laser exposure or temperature cycles was observed. Also, no clear signature of charged excitons and/or trapped excitons is visible in these protected flakes obtained from bulk material of different origins. The temperature-induced broadening of the X$^0$ transition in PL and in reflectivity follows the same linear relationship up to $100$ K, indicating that the PL linewidth contains only small inhomogeneous contributions. \\
\indent We have also revisited magneto-PL measurements with these high quality samples and measured a neutral exciton Land\'e factor of $g_X=-1.7\pm 0.1$ that is smaller in magnitude than the typical value $-4$ reported for other members of the TMDC family and for CVD-grown MoS$_2$ MLs on SiO$_2$ \cite{Li:2014a,srivastava:2015,Macneill:2015a,Arora:2016a,Wang:2015d}. Excitation of this high quality ML with a linearly polarized, non-resonant laser creates a robust coherent superposition of  valley states and a high steady-state linear polarization of the PL (55 $\%$). We show rotation of this coherent superposition of valley states in applied magnetic fields up to 9 T.\\
\indent The well defined optical transitions and negligible defect emission will allow in the future to further explore these stable MoS$_2$ samples, for example, the exciton resonance can be tuned in resonance with optical cavity modes to explore strong coupling between light and matter in microcavities \cite{Liu:2015a,Dufferwiel:2015a} with the added possibility of valley specific optical excitations \cite{Dufferwiel:2016a,Chen:2017a}. Also all time resolved experiments using pulsed laser such as pump-probe spectroscopy, time resolved PL and four-wave mixing \cite{Jakubczyk:2016a,Dey:2016a,Lagarde:2014a,Mai:2014a,yang:2015a} will benefit from these samples with much higher threshold for optical damage as compared to uncapped samples. \\
\indent \textit{Acknowledgements.---} 
We thank ERC Grant No. 306719, ITN Spin-NANO Marie Sklodowska-Curie grant agreement No 676108, ANR MoS2ValleyControl, Programme Investissements d’Avenir ANR-11-IDEX-0002-02, reference ANR-10- LABX-0037-NEXT for financial support. X.M. also acknowledges the Institut Universitaire de France. S.T acknowledges funding from National Science Foundation (DMR-1234567) and CMMI-1561839. K.W. and T.T. acknowledge support from the Elemental Strategy Initiative
conducted by the MEXT, Japan and JSPS KAKENHI Grant Numbers JP26248061,JP15K21722
and JP25106006.\\
\section{Apendix: Experimental details}\label{apendix}
\textit{Optical Spectroscopy.---}
The experiments are carried out at variable temperatures T=4 to 300 K and in magnetic fields up to $\pm$ 9 T in Faraday configuration in a confocal microscope build in a vibration-free, closed cycle cryostat from Attocube. The excitation/detection spot diameter is $\approx1\mu$m, i.e. smaller than the typical ML diameter. The MoS$_2$ ML is excited by continuous wave He-Ne laser (1.96~eV) or at 2.33~eV. The PL signal is dispersed in a spectrometer and detected with a Si-CCD camera. The white light source for reflectivity is a halogen lamp with a stabilized power supply. Faraday effects of the optical set-up in applied fields have been systematically calibrated for plotting the valley coherence rotation angle.\\
\textit{Samples.---}
In addition to bulk material obtained from 2D semiconductors, Ultrahigh purity MoS$_2$ crystals were synthesized using chemical vapor transport (CVT) technique. Precursor MoS$_2$ polycristalline powder was first synthesized by heating and forming a mixture containing stoichiometric amounts of 6N (99.9999 $\%$ purity) sulfur and molybdenum powders (puratronic quality from Alfa Aesar Inc. and American Elements Inc.) at 1020 $\degree$ C for 14 days in an evacuated sealed quartz ampule at pressures less than $10^{-7}$ Torr. In a typical polycristalline MoS$_2$ powder synthesis special care was given to attaining high purity by using chemically cleaned and annealed quartz tubes (to prevent degassing) and mixture was heated from room temperature to 1020 $\degree$ C in 24 h (to prevent explosion). Approximately, 6 grams of MoS$_2$ polycrystals were released together with molybdenum containing high purity (6N) MoCl 5 transport agent (at ca. 4 mg/$\mbox{cm}^3$) in a quartz tube (15.5 mm in diameters and $\sim 18$ cm in length). Here, MoCl 5 sublimates at low temperature and supplies high Mo-based partial pressure to enhance crystal growth. Successful growth is attained, when the source and growth zones are kept at 1065 and 1015 $\degree$ C, respectively, for 12 days.\\
\indent As shown schematically in the inset of Fig.\ref{fig:fig1}a, we have fabricated van der Waals heterostructures by mechanical exfoliation of bulk  MoS$_2$ (CVT-grown or comercially available) and hBN crystals \cite{Taniguchi:2007a}. A first layer of hBN was mechanically exfoliated onto a freshly cleaved SiO$_2$ (90 nm)/Si substrate \cite{Gomez:2014a}. The deposition of the subsequent MoS$_2$ ML and the second hBN capping layer was obtained by repeating this procedure.


\begin{thebibliography}{90}
\bibitem[{Sto()}]{StokesML}
\bibinfo{note}{This makes an exact determination of the Stokes shift between
  reflectivity and PL difficult, as the laser and white light spots need to
  overlap perfectly and have the same diameter. This is of course roughly the
  case in our experiment, and we estimate an upper limit of the Stokes shift of
  5 meV, not excluding that it is close to zero for certain positions on the
  sample.}
\end{thebibliography}

\end{document}